\documentclass{PoS}

\usepackage{amsfonts}
\usepackage{amsmath}
\usepackage{subfigure} 

\graphicspath{{./figs/}}

\title{
  First results for $B_K$ on the ultrafine ($a=0.045$ fm) ensemble
}

\ShortTitle{
  $B_K$ on the ultrafine ensemble
}

\author{\speaker{Taegil Bae}, Yong-Chull Jang,
  Hyung-Jin Kim, Jangho Kim,
  Jongjeong Kim, Kwangwoo Kim, Boram Yoon, Weonjong Lee\\
  Lattice Gauge Theory Research Center, CTP, and FPRD, \\
  Department of Physics and Astronomy,
  Seoul National University, Seoul, 151-747, South Korea \\
  E-mail: \email{wlee@snu.ac.kr}}

\author{Chulwoo Jung \\
  Physics Department, Brookhaven National Laboratory,
  Upton, NY11973, USA \\
  E-mail: \email{chulwoo@bnl.gov}}

\author{Stephen R. Sharpe\\
  Physics Department, University of Washington, Seattle, WA 98195-1560 \\
  E-mail: \email{sharpe@phys.washington.edu}}

\abstract{We present preliminary results for $B_K$ from the
  MILC ultrafine lattices, based on a partial ensemble of 305
  configurations.  We use HYP-smeared improved staggered valence
  quarks.  The analysis is done using fitting forms based on both
  SU(2) and SU(3) staggered chiral perturbation thery.  For the SU(2)
  analysis, we find that the result using the NLO fit function is
  consistent with that from a partial NNLO fit.  For the SU(3)
  analysis, where we have to use partially constrained fits due to the
  number of fit parameters, we find that our two preferred fits
  (``N-BB1'' and ``N-BB2'') are also consistent, both with each other
  and with the results of the SU(2) fits.  These results are
  used in companion proceedings to improve the control over the continuum
  extrapolation.}

\FullConference{
  The XXVIII International Symposium on Lattice Field Theory, Lattice2010\\
  June 14-19, 2010\\
  Villasimius, Italy
}

\begin{document}

\section{Introduction} 
This paper is the last of four proceedings 
\cite{ref:wlee-2010-10,ref:wlee-2010-11,ref:wlee-2010-12}
describing our calculation of $B_K$ using improved staggered fermions.
Here, we present our first results using the MILC ``ultrafine'' ensemble,
with $a=0.045$ fm. This is the finest of four lattice spacings that
we have used, the others being the 
``coarse'' ($a=0.12$ fm), ``fine'' ($a=0.09$ fm)
and ``superfine'' ($a=0.06$ fm) spacings.
Our previous result for $B_K$ was based on these three 
spacings~\cite{ref:wlee-2010-1}.
Having a fourth point closer to the continuum limit
both checks our previous continuum extrapolation and
reduces the error in that extrapolation. 

The parameters for the numerical study are collected in Table
\ref{tab:para}.
As can be seen, we have so far obtained results on only 305 
configurations, less than half of the total available.
Hence, the results are preliminary.
\begin{table}[h!]
\begin{center}
\begin{tabular}{r | l }
\hline
\hline
parameter & value \\
\hline
sea quarks      &  asqtad staggered fermions \\
valence quarks  &  HYP-smeared staggered fermions \\
geometry        &  $64^3 \times 192$ \\
number of confs. &  305 \\
$am_l/am_s$     &  $0.0028/0.014$ \\
$1/a$           &  4517 MeV \\
$\alpha_s$      &  0.2096 for $\mu=1/a$ \\
$am_x$, $am_y$  &  $0.0014 \times n$ ($n=1,2,3,\ldots,10$) \\ 
\hline
\hline
\end{tabular}
\end{center}
\caption{Parameters for the numerical study on the ultrafine (U1) ensemble.
  $m_l$ is the light sea quark mass, $m_s$ the strange sea quark mass,
  $m_x$ the light valence quark mass, and $m_y$ the strange valence quark 
  mass.}
\label{tab:para}
\end{table}

\section{Extracting $B_K$}
\begin{figure}[tbhp]
\centering
\includegraphics[width=0.7\textwidth]
{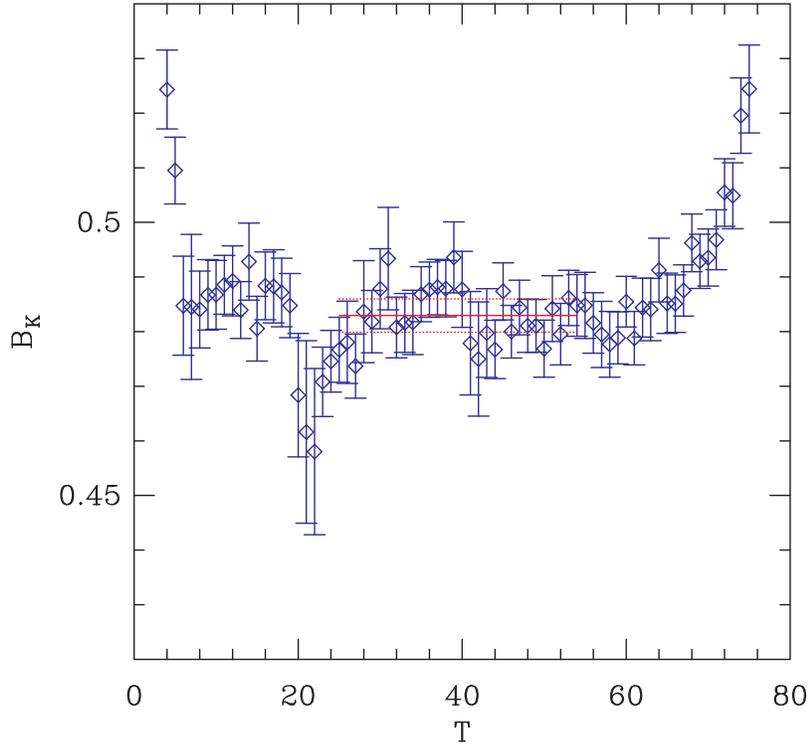}
\caption{$B_K$ versus $T$ on the ultrafine ensemble, obtained using
  one-loop matching with renormalization scale $\mu=1/a$. 
  Valence masses are $am_x=am_y=0.007$, corresponding to a kaon with
  approximately the physical mass.}
\label{fig:bk-t}
\end{figure}

We calculate $B_K$ using the methods described in Ref.~\cite{ref:wlee-2010-1}.
We place the U(1) noise wall-sources at $t=0$ and $t=80$.
These sources couple only to the Goldstone-taste pion ($\xi_5$).
In Fig.~\ref{fig:bk-t}, we show an example of
our results for the ratio of matrix elements which equals $B_K$
when the operator (placed at time $T$) is far from both sources.
We choose the fitting range such that 
the excited states with
the same quantum numbers as the Goldstone pion mode 
do not make significant contributions,
as determined from wall-source to current correlators.
In this case, we choose the fitting range to be $ 25 \le t \le 54$,
and fit to a constant, as in the example
shown in Fig.~\ref{fig:bk-t}.

\section{SU(2) SChPT analysis}
Our most reliable method of extrapolating $B_K$ to the physical quark masses
is based on SU(2) staggered chiral perturbation theory (SChPT). 
The resulting fit forms and a detailed description of our fitting method
are given in Ref.~\cite{ref:wlee-2010-1}.
In brief, the fits are done in two steps.
\begin{enumerate}
\item
In the ``X-fit'' we extrapolate $m_x \to m_d^{\rm phys}$,
with $m_y$ fixed,
while at the same time using the SChPT fit form 
to remove taste-breaking lattice artifacts, and to set
$m_\ell \to m_\ell^{\rm phys}$ in the chiral logarithms.
The fit function takes the form~\cite{ref:wlee-2010-1}
\begin{eqnarray}
  f_\text{th} &=& d_0 F_0 + d_1 \frac{X_P}{\Lambda_\chi^2}  + d_2
  \frac{X_P^2}{\Lambda_\chi^4} \,,
\label{eq:fth}
\end{eqnarray}
where $X_P$ is the mass of the pion composed of valence quarks
of mass $m_x$, and $F_0 = 1 + \text{chiral logs}$.
The chiral logarithms have a known form in terms of measurable pion masses.
The coefficient $d_2$ multiplies an analytic next-to-next-to-leading order (NNLO)
term. The coefficients $d_0-d_2$ are expected to have $O(1)$ magnitudes.
\item
In the ``Y-fit'', we extrapolate the results of
the X-fits to $m_y = m_s^{\rm phys}$, using an analytic
fit function. A linear Y-fit appears sufficient, and we use this
for our central value.
\end{enumerate}
\begin{figure}[tbhp]
\centering
\includegraphics[width=0.49\textwidth]
{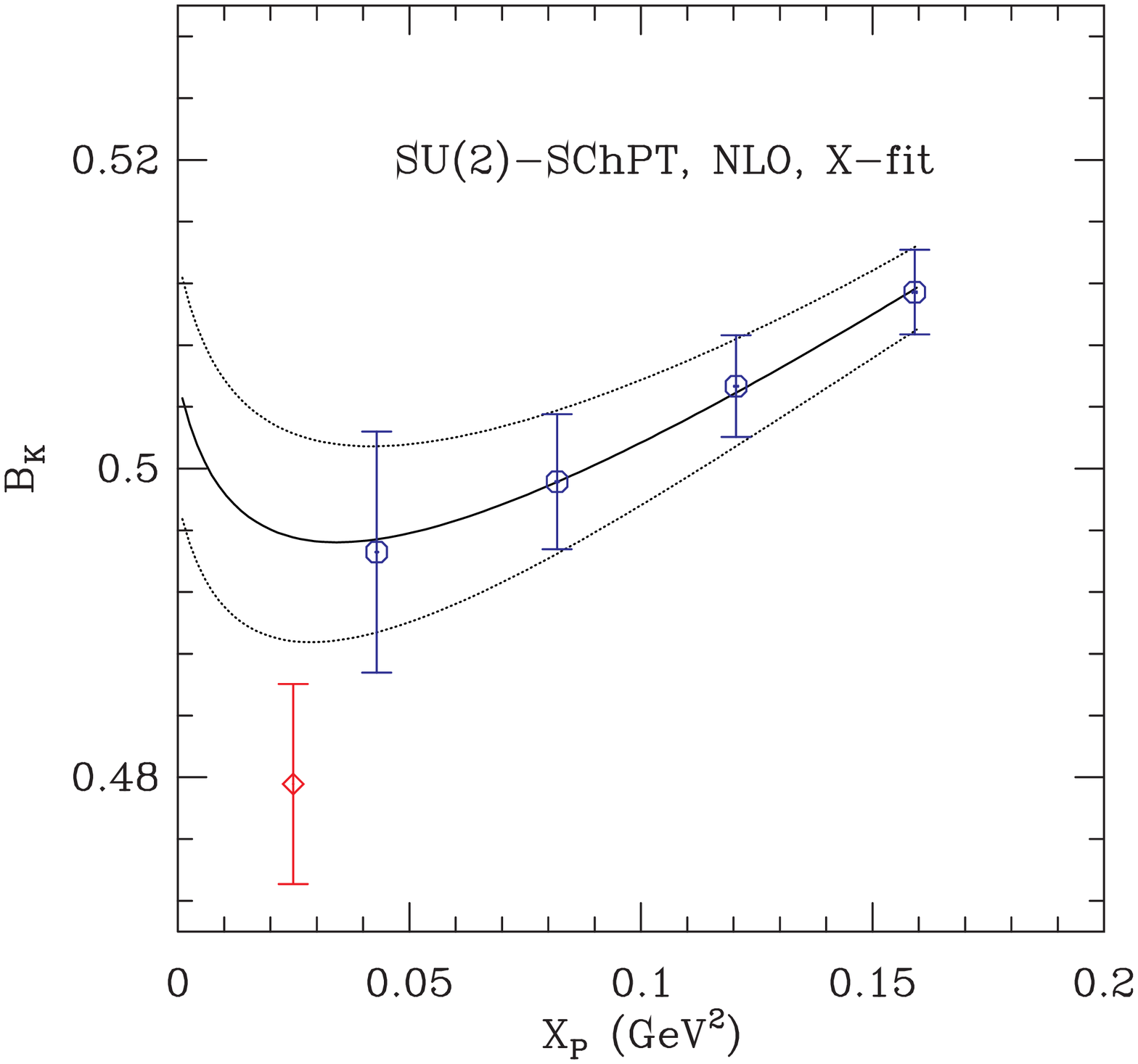}
\includegraphics[width=0.49\textwidth]
{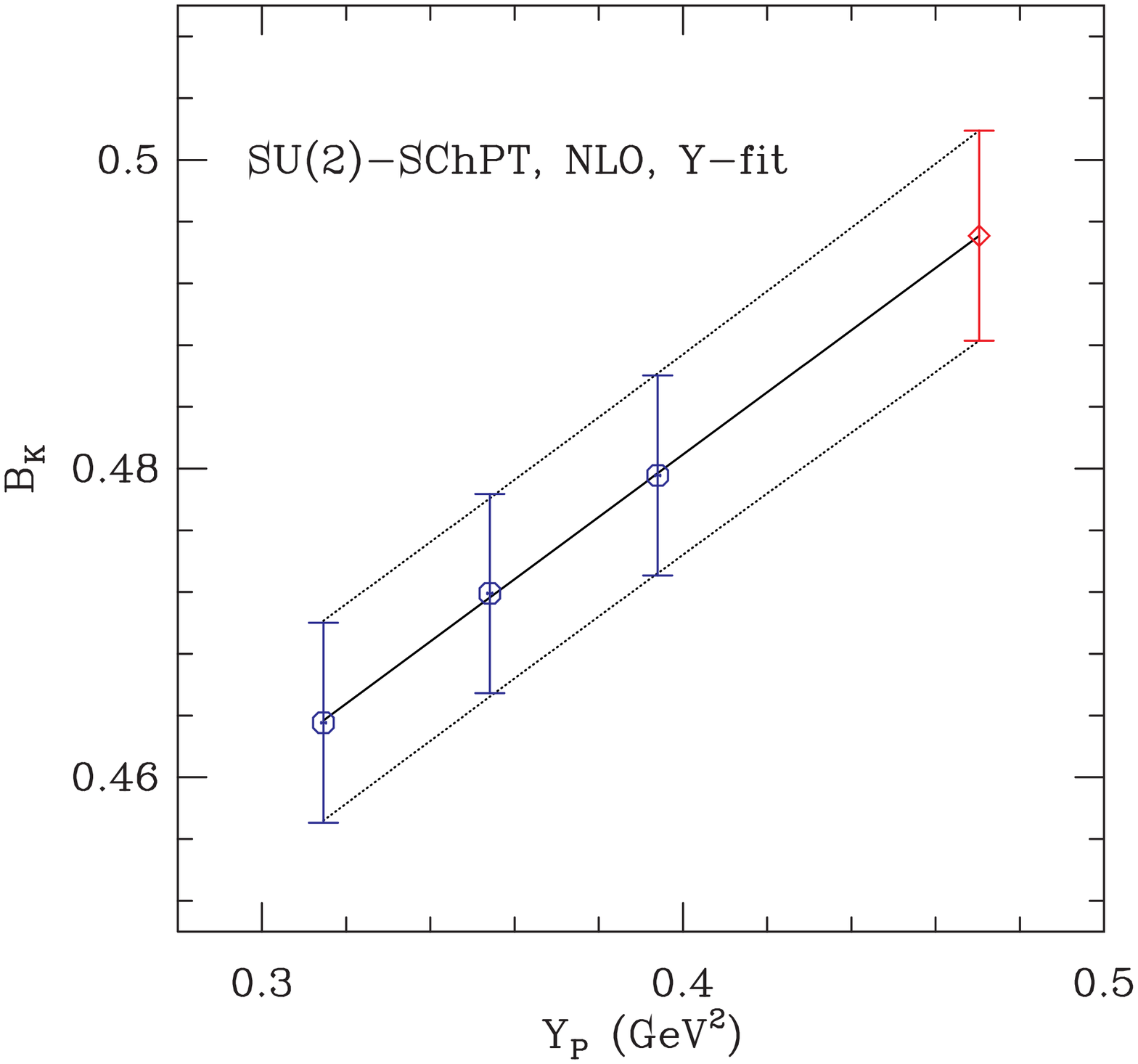}
\caption{$B_K$ versus $X_P$ (mass of the pion composed of valence
  quarks of mass $m_x$) and corresponding 4X-NLO fit (left panel) and
  $B_K$ versus $Y_P$ (mass of pion composed of quarks of mass $m_y$)
  along with the 3Y linear fit (right panel).  $B_K(\mu=1/a)$ is
  obtained using one-loop matching.  In the left panel, $a m_y=0.014$,
  and the red point is the result of setting $m_x=m_d^{\rm phys}$,
  $m_\ell=m_\ell^{\rm phys}$ and removing taste breaking.  }
\label{fig:su2-4x3y-nlo}
\end{figure}
In Fig.~\ref{fig:su2-4x3y-nlo}, we show an example of both
X- and Y-fits.
We use our lightest four values of $m_x$ in the X-fits and
our heaviest three values of $m_y$ in the Y-fits.
Furthermore, the fit function for the X-fit is of NLO.
Thus the fit is labeled 4X3Y-NLO.
The corresponding plots for NNLO X-fits, in which
we add a single analytic NNLO term~\cite{ref:wlee-2010-1},
are shown in Fig.~\ref{fig:su2-4x3y-nnlo}.
Illustrative parameters from these fits are collected in
Table~\ref{tab:X-fit:U1}. Note that, since we use an
uncorrelated $\chi^2/{\rm d.o.f}$, we expect values
much smaller than unity if the fit is good.
\begin{figure}[tbhp]
\centering
\includegraphics[width=0.49\textwidth]
{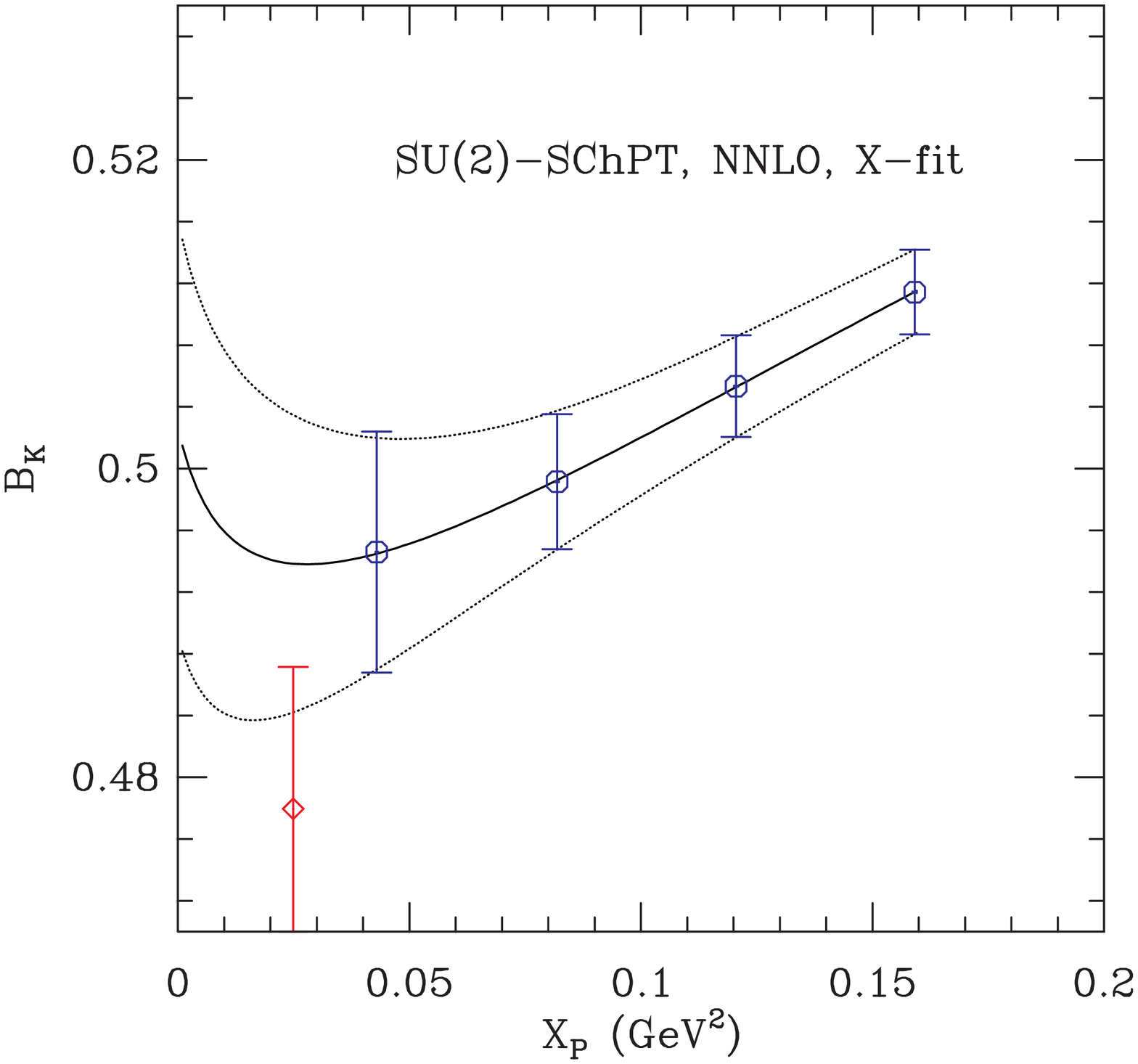}
\includegraphics[width=0.49\textwidth]
{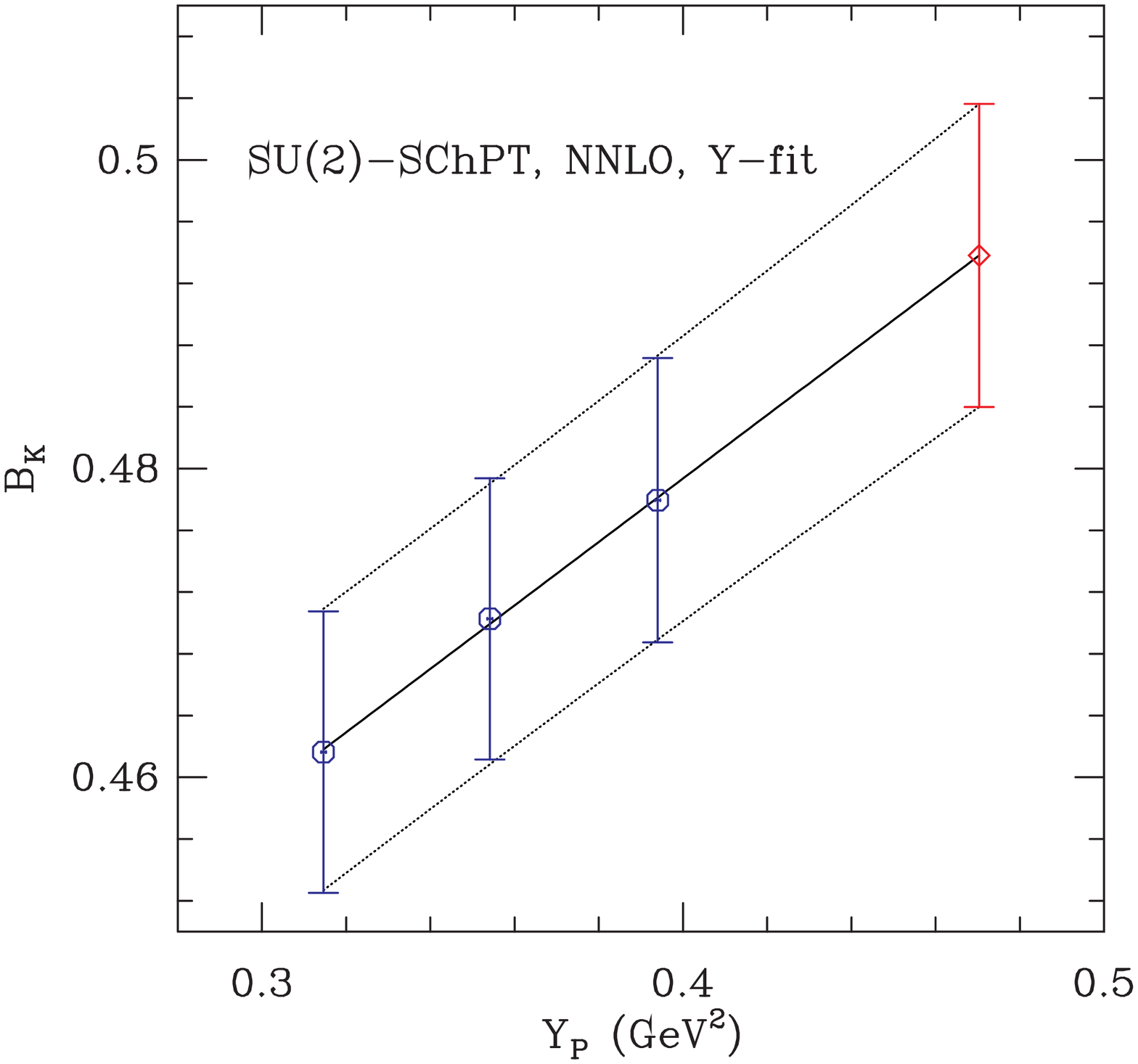}
\caption{As for Fig.~\protect\ref{fig:su2-4x3y-nlo},
but using NNLO X-fits. }
\label{fig:su2-4x3y-nnlo}
\end{figure}
%

%
%
%
\begin{table}[h!]
\begin{center}
\begin{tabular}{c | c | c | c | c || c }
\hline
\hline
fit & $d_0$ & $d_1$ & $d_2$ & $\chi^2/\text{d.o.f}$ & $B_K(\mu=1/a)$ \\
\hline
4X3Y-NLO  & 0.4702(73)  & 0.155(40) &     ---     & 0.017(66)  & 0.4951(68) \\
4X3Y-NNLO & 0.4973(126) & 0.22(17)  & $-$0.26(57) & 0.0012(99) & 0.4938(98) \\
\hline
\hline
\end{tabular}
\end{center}
\caption{X-fit parameters for the 4X-NLO and 4X-NNLO fits shown in the left
panels of Figs.~\protect\ref{fig:su2-4x3y-nlo}
and \protect\ref{fig:su2-4x3y-nnlo}, 
and results for $B_K$ after the Y-fits shown in the right panels.}
\label{tab:X-fit:U1}
\end{table}

We can see from the figures and the Table that the difference
between values for $B_K$ resulting from these fits is very small.
We also see that, although $d_0$ is well determined, $d_1$ and $d_2$ are not.
The poor determination of $d_1$ and $d_2$ has, however, little impact
on our extrapolated value.

An important feature of the fitting function is the
presence of chiral logarithms, which lead to the curvature for small $X_P$.
While our data is consistent with this curvature, it 
is very small in the region of our points, and our data itself provides no direct
evidence for the presence of chiral logarithms.

Finally, we note that the convergence of SU(2) ChPT is 
satisfactory for all points included in the fits. This can be seen, for example,
by the closeness of $d_0$ to the values of $B_K$ to which we fit.

\section{SU(3) Analysis}

The SU(3) fits have the advantage of using all our data points
(55 mass pairs), but two disadvantages.
These are that the convergence of SU(3) ChPT is
questionable for many of our points, and
that the NLO SU(3) SChPT fit forms are much more involved.
We sketch the situation here, and refer to 
Ref.~\cite{ref:wlee-2010-1} for details.

Compared to the SU(2) fits, we have to make two simplifications
in order to obtain stable fits. First, we need to lump together classes of
fit functions which have similar functional forms, using only one
function as representative. Second, we need to use constrained
(Bayesian) fitting---parameters associated with taste-breaking
are constrained to lie in the range of values expected from 
SChPT power counting.
With these simplifications we were able to obtain good fits
on the coarse, fine and superfine ensembles~\cite{ref:wlee-2010-1}.
We found, however, that different assumptions about the constraints
led to significant variation in the final answer for $B_K$,
and this gave rise to a large systematic error.
The particular value for this error quoted in \cite{ref:wlee-2010-1},
which was 5.3\%,
came from the analysis on the coarse ensemble C3.
Thus it is interesting to see whether the variation between fits
is reduced on the ultrafine lattices.
We would expect significant reduction because the offending terms
in the fit functions are proportional to either $a^2$ or
$\alpha_s(1/a)^2$.
\begin{figure}[t!]
\centering
\includegraphics[width=0.49\textwidth]
{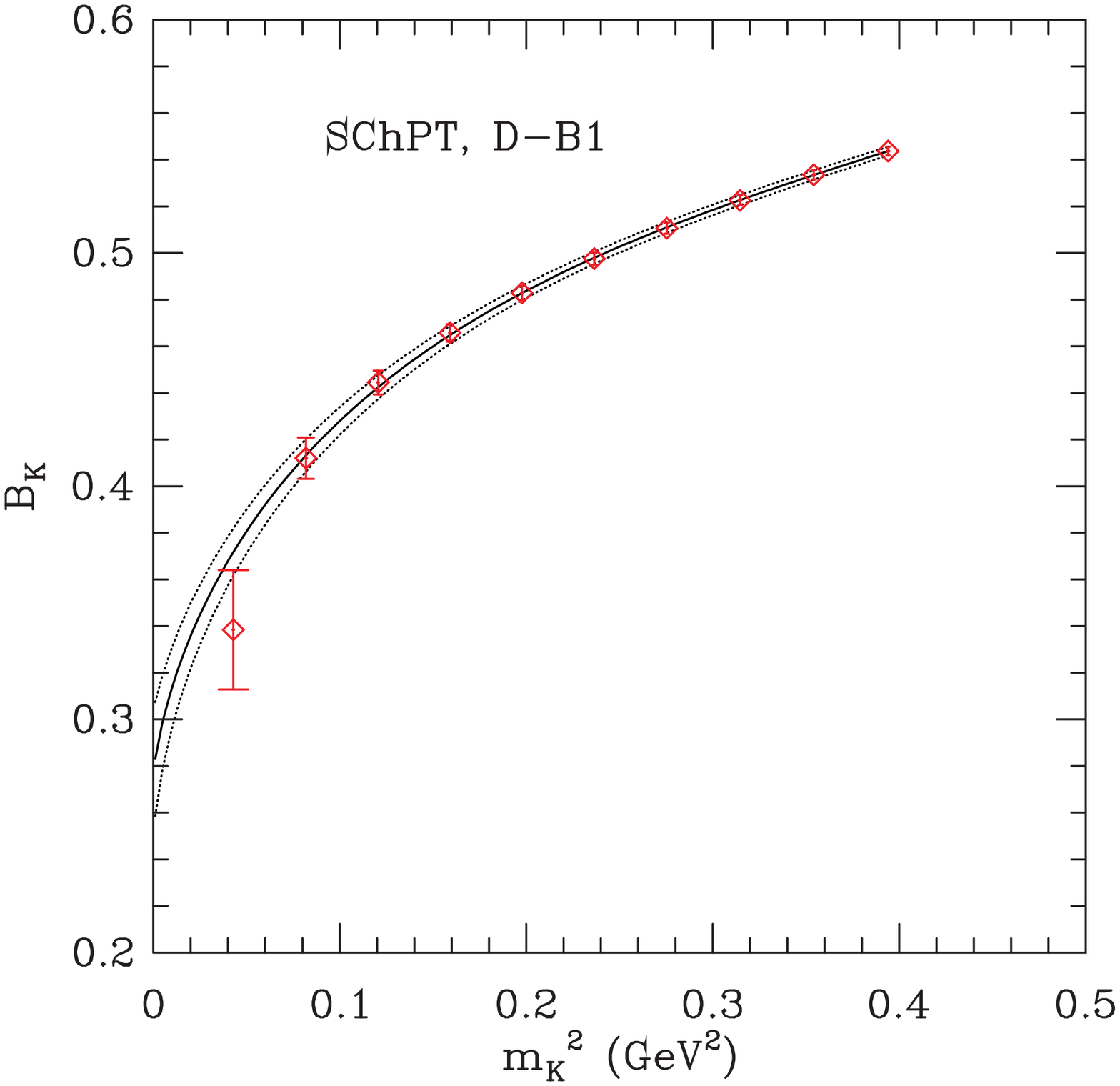}
\includegraphics[width=0.49\textwidth]
{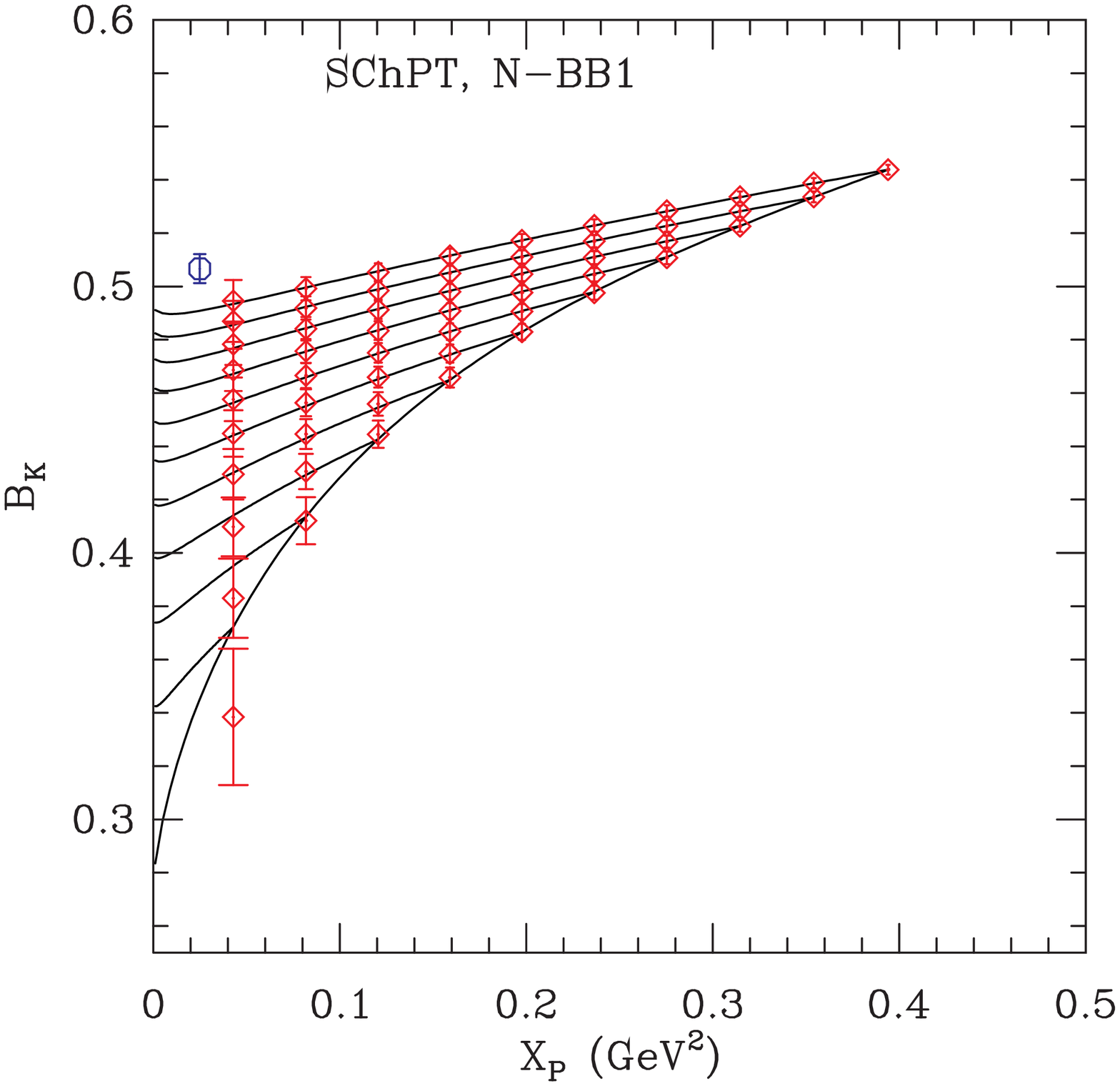}
\caption{ One-loop matched $B_K$ 
versus $X_P$ on the U1 lattice for the D-B1 fit (left)
  and for the N-BB1 fit (right).  The fits are described in the text
and, in more detail, in Ref.~\protect\cite{ref:wlee-2010-1}.
The blue octagon in the right-hand plot
shows the result after extrapolation to physical quark masses
and after removing lattice artifacts.}
\label{fig:su3-nbb1}
\end{figure}
In Fig.~\ref{fig:su3-nbb1}, we show the results of two fits.
That shown on the left is a ``D-B1'' fit,
in which we use only degenerate kaons, and constrain
the representative lattice artifact term based on the
assumption that it is proportional to $a^2$.
The fit in the right panel is a ``D-BB1'' fit to the
entire data set, with two Bayesian constraints.
The first constraint 
is that the results (including errors) of the D-B1 fit are used
to constrain the parameters that contribute for degenerate kaons.
The second is that additional representative lattice artifact
terms are constrained based on the assumption that they
are proportional to $a^2$.

\begin{figure}[t!]
\centering
\includegraphics[width=0.49\textwidth]
{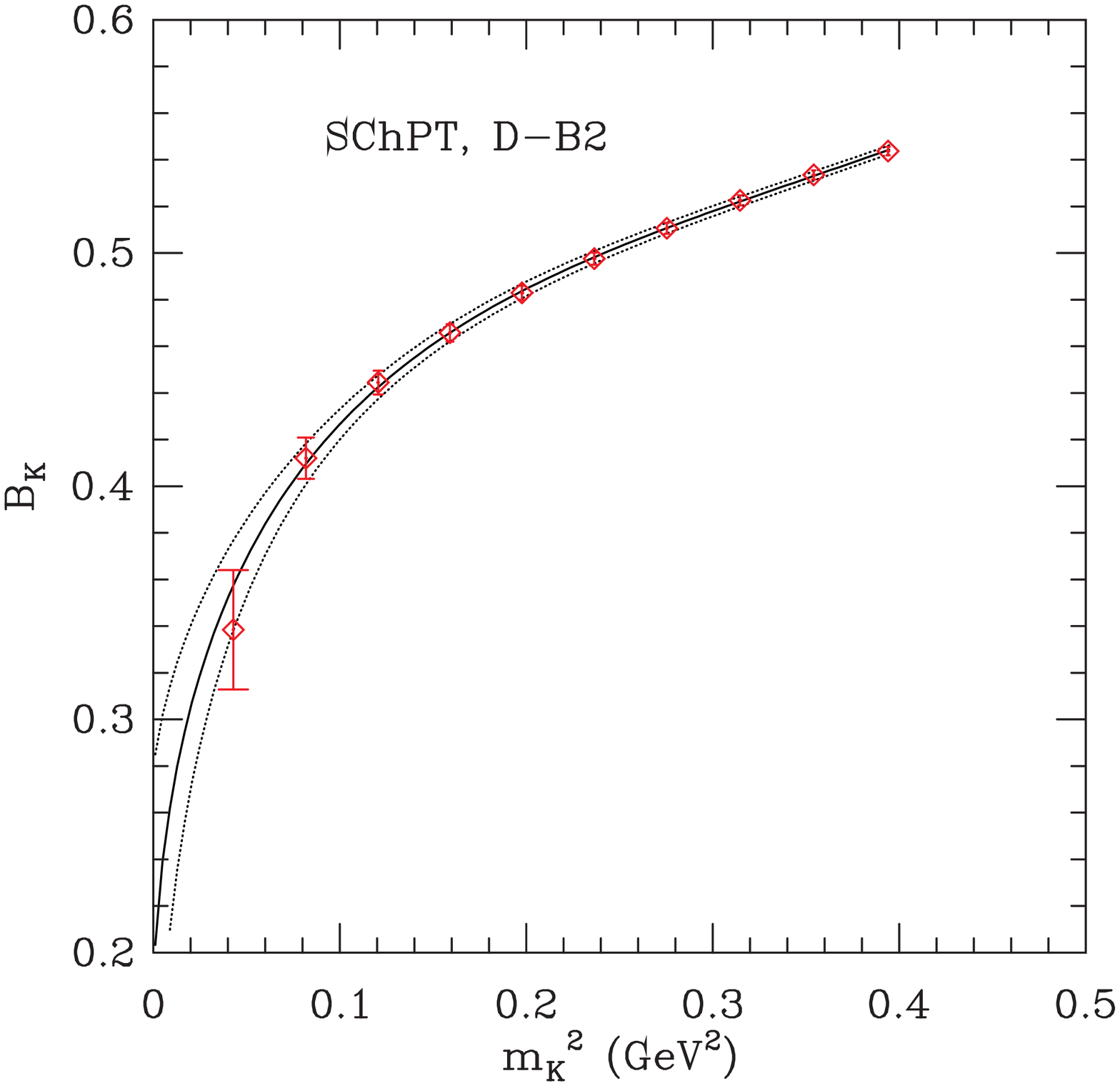}
\includegraphics[width=0.49\textwidth]
{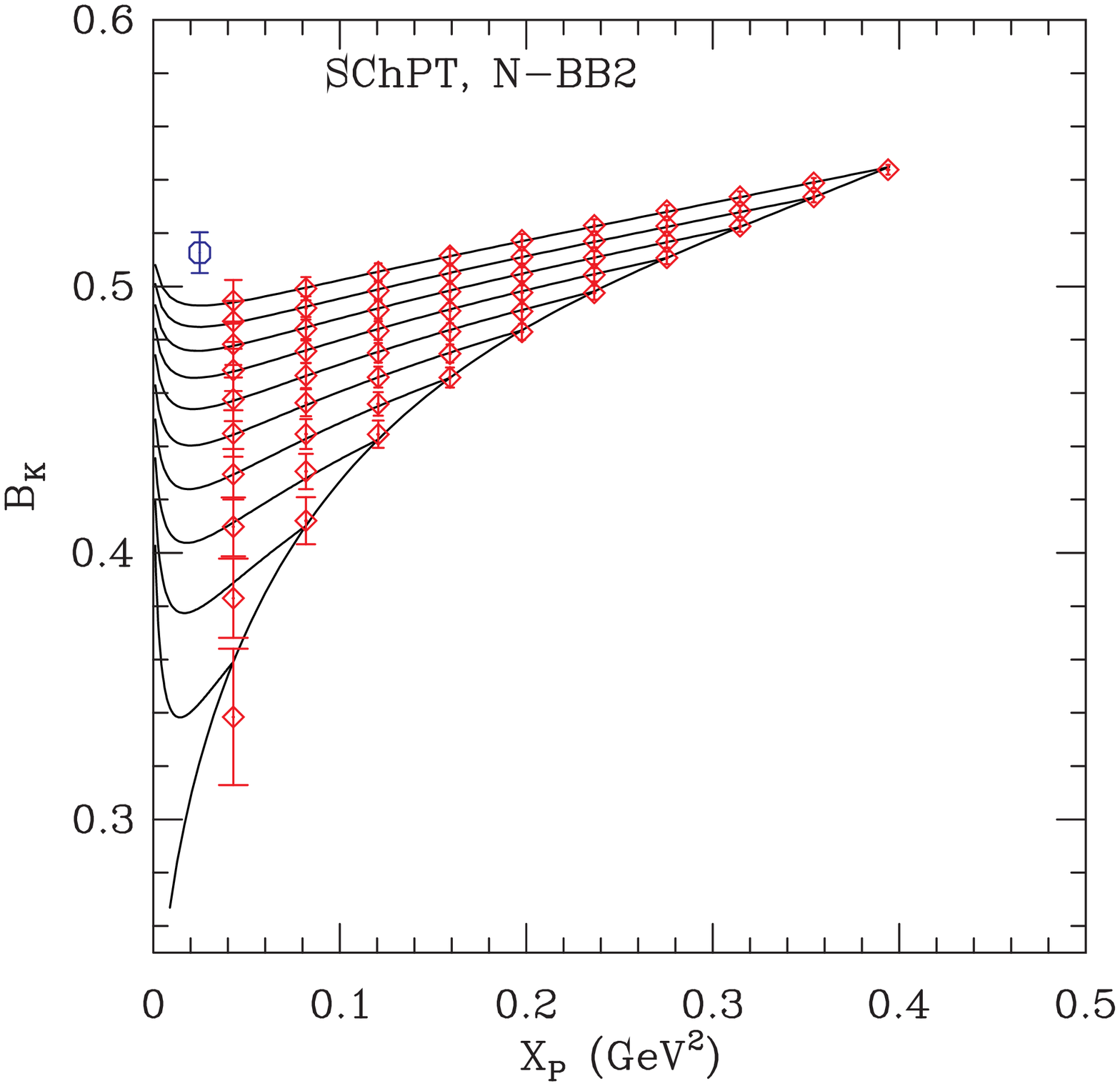}
\caption{As for Fig.~\protect\ref{fig:su3-nbb1},
except using D-B2 (left) and N-BB2 (right) fits.}
\label{fig:su3-nbb2}
\end{figure}

In Fig.~\ref{fig:su3-nbb2}, we show the $B_K$ results of two different
fits: D-B2 on the left and N-BB2 on the right.
These differ from D-B1 and N-BB1 in that lattice artifact terms
are presumed to be proportional to $\alpha_s^2$ rather than $a^2$.
The results for $B_K(1/a)$, after removal of taste-breaking
from the chiral logarithms, and after extrapolation to physical
quark masses, are given in Table~\ref{tab:N-BB1,N-BB2}.

\begin{table}[htbp]
\begin{center}
\begin{tabular}{c | c | c }
\hline
\hline
fit type & $\chi^2/\text{d.o.f}$ & $B_K(\mu=1/a)$\\
\hline
N-BB1 &  0.075(81) &  0.5067(55) \\
N-BB2 &  0.052(38) &  0.5127(76) \\
\hline
\hline
\end{tabular}
\end{center}
\caption{Results for $B_K$ and fit quality from N-BB1 and N-BB2 fits 
on the U1 ensemble.}
\label{tab:N-BB1,N-BB2}
\end{table}

We see that the difference between N-BB1 and N-BB2 fits has
been reduced to 1.2\%, compared to the 5.3\% found on the C3 ensemble.
We interpret this improvement as being due to the reduction in
the size of taste-breaking effects. As Table~\ref{tab:a^2,alpha^2}
shows, if these effects are dominantly discretization errors,
the expected reduction in their size is $\sim 7.5$, while
if they are dominantly truncation errors the reduction is $\sim 2.5$.
Thus, as far as the need for Bayesian constraints goes,
the move to smaller lattice spacings improves
the stability of the SU(3) fitting.
The only caveat to this statement is that a similar reduction is
not observed on the S1 ensemble~\cite{ref:wlee-2010-1}.

\begin{table}[htbp]
\begin{center}
\begin{tabular}{c | c | c }
\hline
\hline
parameter & C3 &  U1 \\
\hline
$a^2$ (fm${}^2$)      & 0.0141 & 0.0019 \\
$\alpha_s^2(\mu=1/a)$ & 0.1080 & 0.0439 \\
\hline
\hline
\end{tabular}
\end{center}
\caption{$a^2$ and $\alpha_s^2$ for the C3 and U1 lattices.}
\label{tab:a^2,alpha^2}
\end{table}

\section{Conclusion}

As one can see from Tables~\ref{tab:X-fit:U1} and \ref{tab:N-BB1,N-BB2}, 
the results from the SU(2) fits lie below those from the SU(3) fits.
This difference is not, however, statistically significant.
For example, using
our preferred 4X3Y-NNLO and N-BB1 fits, the difference is only $1.2\sigma$.
Although the error on the SU(3) N-BB1 fit is nominally smaller than
that on the SU(2) 4X3Y-NNLO fit,  we think that the latter fit is more reliable,
as discussed above and in Ref.~\cite{ref:wlee-2010-1}.

Although our results use somewhat less than half of the 883
ultrafine configurations that we intend to analyze, we can
already see that fitting simplifies as we approach the
continuum limit, and systematic errors are correspondingly reduced. 
This is particularly true of the SU(3) fitting.
However, one should keep in mind that one does not expect the
chiral convergence of SU(3) fitting to improve on finer lattices.

The data presented here is incorporated into continuum extrapolations
in two of the companion proceedings
\cite{ref:wlee-2010-10,ref:wlee-2010-11}.

\section{Acknowledgments}
C.~Jung is supported by the US DOE under contract DE-AC02-98CH10886.
The research of W.~Lee is supported by the Creative Research
Initiatives Program (3348-20090015) of the NRF grant funded by the
Korean government (MEST). 
The work of S.~Sharpe is supported in part by the US DOE grant
no.~DE-FG02-96ER40956.
Computations were carried out in part on QCDOC computing facilities of
the USQCD Collaboration at Brookhaven National Lab. The USQCD
Collaboration are funded by the Office of Science of the
U.S. Department of Energy.

\end{document}